\documentclass[superscriptaddress,twocolumn,showpacs,reprint,amsmath,amssymb]{revtex4}

\usepackage{graphicx}
\usepackage{epstopdf}
\usepackage{bm}
\usepackage{amsmath}

\newcommand{\tn}{\textnormal}

\begin{document}

\title{Quasiparticles dynamics in high-temperature superconductors far from equilibrium: an indication of pairing amplitude without phase coherence}

\author{C. Piovera}
\affiliation{Laboratoire des Solides Irradi\'es, CNRS UMR 7642, CEA-DSM-IRAMIS, Ecole polytechnique, Universit\'e Paris-Saclay, 91128 PALAISEAU cedex, France}

\author{Z. Zhang}
\affiliation{Institut de Min\'eralogie, de Physique des Mat\'eriaux, et de Cosmochimie (IMPMC), Sorbonne Universit\'es - UPMC Univ Paris 06,  case 115, 4, place Jussieu, 75252 Paris cedex 05, France }

\author{M. d'Astuto}
\affiliation{Institut de Min\'eralogie, de Physique des Mat\'eriaux, et de Cosmochimie (IMPMC), Sorbonne Universit\'es - UPMC Univ Paris 06,  case 115, 4, place Jussieu, 75252 Paris cedex 05, France }

\author{A. Taleb-Ibrahimi}
\affiliation{Synchrotron SOLEIL, L’Orme des Merisiers, Saint-Aubin-BP 48, F-91192 Gif sur Yvette, France}

\author{E. Papalazarou}
\affiliation{Laboratoire de Physique des Solides, CNRS-UMR 8502, Universit\'e Paris-Sud, FR-91405 Orsay, France}

\author{M. Marsi}
\affiliation{Laboratoire de Physique des Solides, CNRS-UMR 8502, Universit\'e Paris-Sud, FR-91405 Orsay, France}

\author{Z. Z. Li}
\affiliation{Laboratoire de Physique des Solides, CNRS-UMR 8502, Universit\'e Paris-Sud, FR-91405 Orsay, France}

\author{H. Raffy}
\affiliation{Laboratoire de Physique des Solides, CNRS-UMR 8502, Universit\'e Paris-Sud, FR-91405 Orsay, France}

\author{L. Perfetti}
\affiliation{Laboratoire des Solides Irradi\'es, CNRS UMR 7642, CEA-DSM-IRAMIS, Ecole polytechnique, Universit\'e Paris-Saclay, 91128 PALAISEAU cedex, France}

\begin{abstract}

We perform time resolved photoelectron spectroscopy measurements of optimally doped $\tn{Bi}_2\tn{Sr}_2\tn{CaCu}_2\tn{O}_{8+\delta}$ (Bi-2212) and  $\tn{Bi}_2\tn{Sr}_{2-x}\tn{La}_{x}\tn{Cu}\tn{O}_{6+\delta}$ (Bi-2201). The electrons dynamics show that  inelastic scattering by nodal quasiparticles decreases when the temperature is lowered below the critical value of the superconducting phase transition. This drop of electronic dissipation is astonishingly robust and survives to photoexcitation densities much larger than the value sustained by long-range superconductivity. The unconventional behaviour of quasiparticle scattering is ascribed to superconducting correlations extending on a length scale comparable to the inelastic path. Our measurements indicate that strongly driven superconductors enter in a regime without phase coherence but finite pairing amplitude. The latter vanishes near to the critical temperature and has no evident link with the pseudogap observed by Angle Resolved Photoelectron Spectroscopy (ARPES).

\end{abstract}

\maketitle

The equilibrium properties of cuprates superconductors have been characterized by an impressive number of different techniques. The normal phase of these compounds displays an antinodal pseudogap whose origin is still debated \cite{Norman}. In the superconducting phase, quasiparticles are well defined also at the antinodes and generate a single particle gap with $d$-wave simmetry \cite{Makoto,Campuzano}. Conversely to conventional superconductors, the cuprates display a layered structure and a pairing interaction extending over few lattice sites. Therefore precursor effects of the superconducting condensate can be observed slightly above the transition temperature $T_c$ \cite{Alloul,Bergeal,Ong,Perfetti0}. In this critical region, the amplitude fluctuations of the order parameter account for most of the experimental results \cite{Alloul,Bergeal,Perfetti0}.

Whether incoherent Cooper pairs exist far from equilibrium conditions is at the focus of our current research activity. Many pump-probe experiments have already monitored the dynamics of the condensate in cuprates.  It is established that superconductivity recovers in several picoseconds and that such timescale becomes fluence dependent in the low excitation regime \cite{Demsar,Segre}. These results have been confirmed by monitoring the inductive response of supercurrents \cite{Kaindl0} and the gapped spectrum of the single particle excitations \cite{Lanzara0,Lanzara1}. Phenomenological rate equations could successfully account for the recovery of superconductivity in the weak perturbation regime \cite{Kaindl0,Kabanov}. Nonetheless, the photoexcited state generated by intense optical pulses is still poorly understood. Apparently, the dynamics of transient reflectivity display a fast relaxation channel only if the sample is above $T_c$ or if it is strongly photoexcited \cite{Giannetti0, Mihailovic}. Below $T_c$, the lack of a fast relaxation suggests that superconducting correlations inhibit dissipation of quasiparticles \cite{Giannetti0}. Time ad Angle Resolved PhotoElectron Spectroscopy (tr-ARPES) confirmed the appearance of a fast dynamics only in the high excitation regime of the superconducting phase \cite{Bovensiepen0}.

In this work, we show that a detailed (tr-ARPES) analysis of the quasiparticles dynamics provides deep insights on the photoexcited state. The employed pumping fluence is always above the largest value sustained by long-range superconductivity. Despite it, the inelastic scattering of photoexcited quasiparticles displays a downturn below $T_c$. We ascribe such unusual finding to the persistence of short range superconducting correlations up to large photoexcitation densities. It follows that strongly driven condensates are in a transient state with no phase coherence but strong pairing amplitude. This state is qualitatively different from the fluctuating superconductor in equilibrium conditions. In the latter case, our measurements confirm that the pairing amplitude vanish when the temperature is raised slightly above $T_c$.

\section{method}

\begin{figure} \begin{center}
\includegraphics[width=1\columnwidth]{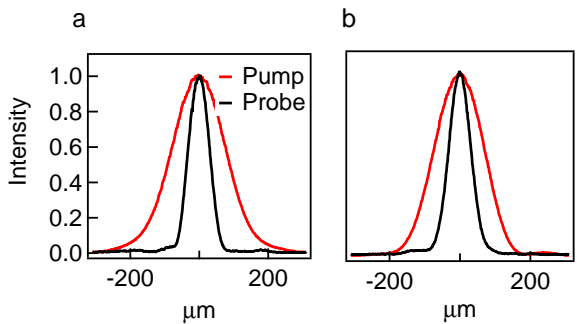}

\caption{ Horizontal a) and vertical b) profiles of the pump and probe beam on the camera. The best spatial overlap on the camera results in the highest pump-probe signal.}\label{Fig1}
\end{center}
\end{figure}

We perform tr-ARPES on optimally doped $\tn{Bi}_2\tn{Sr}_2\tn{CaCu}_2\tn{O}_{8+\delta}$ (Bi-2212) single crystals ( $T_c=91$ K ) and thin film of optimally doped $\tn{Bi}_2\tn{Sr}_{2-x}\tn{La}_{x}\tn{Cu}\tn{O}_{6+\delta}$ Bi-2201 ( $T_c=28$ K ). The samples are mounted on a cryogenic manipulator and are cleaved at the base pressure of 6x10$^{-11}$ mbar.  Time-resolved measurements are carried out at 250 KHz in a pump-probe scheme: short and intense 40 fs laser pulses at a central energy of 1.55 eV drive the system far from equilibrium whereas the photoelectrons are emitted by time delayed pulses at 6.28 eV.  Pump and probe beams are focused almost collinearly on the sample and have a cross correlation with FWHM of about 80 fs \cite{Faure}. The fluence of the pumping pulses has been carefully measured by imaging the focal point of the laser beams on an external camera. Figure \ref{Fig1} shows the horizontal and vertical profile of the pump and probe beam when the spatial overlap and the pump-probe signal are maximal. We measure the average power just before the entrance in the UHV chamber and we weight the pump profile with the probe one. By these means, we can precisely estimate the average fluence incident on the probed area of the sample. We set the probe beam polarization along the nodal plane of the crystal in order to maximize the photoelectron signal generated by the quasiparticles. Photoelectron spectra are acquired with an angular resolution better than 0.1$^\circ$ and an energy resolution of 70  meV. The typical probing depth of emitted photoelectrons is few nanometers whereas the optical penetration of the pump beam is roughly 150 nm. As a consequence, our experiment probes a region at the surface of the sample with nearly uniform excitation density. In order to avoid thermal heating, the repetition rate has been reduced to 100 KHz for pumping fluence above 150 $\mu$Joule/cm$^2$. The absence of a photoinduced signal at negative pump-probe delay guaranties that the average heating of the surface is always negligible.

\section{Quasiparticle relaxation above the transition temperature}

\begin{figure} \begin{center}
\includegraphics[width=1\columnwidth]{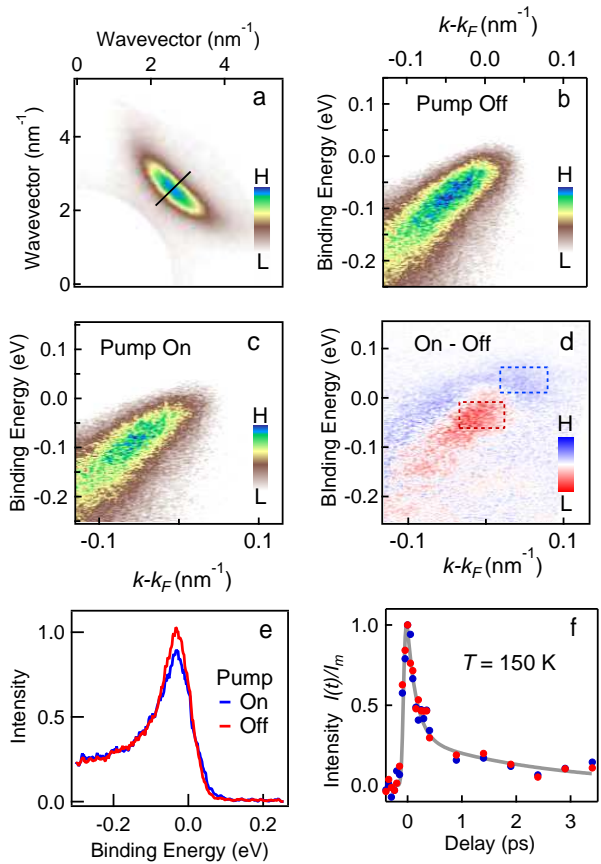}

\caption{The data of this figure have been acquired on optimally doped Bi-2212 ($T_c=91$ K) at $T=150$ K. a) Map of the photoelectron intensity integrated in a small energy interval centered around the Fermi level. The black line visualizes the cut along the nodal direction where we perform tr-ARPES measurements. b) Photoelectron intensity map showing the quasiparticle dispersion along the nodal direction without the pump pulse. c) Photoelectron intensity map aquired with pump pulse at delay time of 50 fs. d) Pump-on minus pump-off intensity map at delay time $t=50$ fs. The dashed lines indicate the areas where the signal of photoexcited electrons (blue) and photoexcited holes (red) have been integrated. e) Energy distribution curves extracted at the Fermi wavevector and aquired with (blue) or without (red) pump pulse. f) Temporal evolution of the photoexcited electrons (blue) and holes (red) acquired with fluence of $60 \mu$J/cm$^2$. The solid line is a bi-exponential fit convoluted with our cross correlation.}\label{Fig2}
\end{center}
\end{figure}

As sketched in Fig. \ref{Fig2}(a), the tr-ARPES spectra are measured in the nodal direction of the first Brillouin zone. Figure \ref{Fig2}(b) shows the quasiparticle dispersion of optimally doped Bi-2212 measured at $T=150$ K in equilibrium condition, i.e without pump irradiation. We show in Fig. \ref{Fig2}(c) the photoelectron intensity map acquired at 50 fs after the arrival of a pump-pulse carring 60 $\mu$Joules/cm$^2$. As shown in Fig. \ref{Fig2}(d), effect of the pump excitation can be enlightened by the subtraction between the photoelectron intensity map acquired at positive delay and the one acquired without pump pulse. Here two major effects can be resolved: (i) a transfer of spectral weight from below the Fermi level (red in false colors) to above it (in blue) \cite{Lanzara0} and (ii) a rigid band shift joined to the photoinduced broadening of the quasiparticle peak \cite{Bovensiepen1}. Effect (i) is also visible in the energy distribution curves extracted at the Fermi wavevector and plot in Fig. \ref{Fig2}(e). The additional shift (ii) generates an area of intensity gain below the Fermi level (blue in Fig. \ref{Fig2}(d)). Finally the photoinduced broadening must be extracted from an analysis of the momentum distribution curves. In the following we focus only on (i), i.e. on the recombination processes that drive the system back to the equilibrium.

\begin{figure} \begin{center}
\includegraphics[width=1\columnwidth]{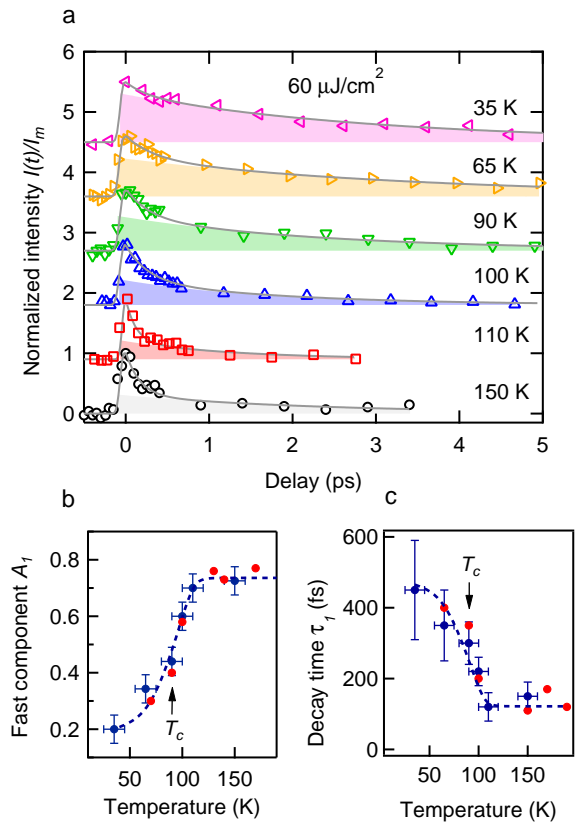}

\caption{The data of this figure have been acquired on optimally doped Bi-2212 ($T_c=91$ K) with pumping fluence $60 \mu$J/cm$^2$. a) Dynamics of photoexcited quasiparticles acquired at different temperatures. Solid lines are bi-exponential fits $A_1\exp(-t/\tau_1)+A_2\exp(-t/\tau_2)$. The underneath colored areas stand for the slow component $A_2\exp(-t/\tau_2)$. The curves have been shifted by an arbitrary offset for better clarity. b) Relative weight of the fast component $A_1$ as a function of temperature. c) Fast decay time $\tau_1$ as a function of temperature. Blue and red marks in panel (b,c) are two different cleaves while the dashed line is a guide to the eye.}\label{Fig3}
\end{center}
\end{figure}

We track the transient dynamics by integrating the intensity of the differential image in an area just above  and just below the Fermi level (see Fig. \ref{Fig2} (d)). Figure 2 (f) shows the obtained intensity $I(t)$ of photoexcited electrons (blue) and photoexcited holes (red) normalized to the maximal value $I_{m}$. We fit $I(t)/I_{m}$ by a bi-exponential function $A_1\exp(−t/\tau_1)+A_2\exp(−t/\tau_2)$ convoluted with a Gaussian distribution of 80 fs FWHM. In all measured cases the dynamics of photoexcited electrons and photoexcited holes are identical within the errors bars. The relaxation takes place on two distinct timescales: a faster one with a decay time $\tau_1=150$ fs and a slower one with $\tau_2=2.5$ ps \cite{Perfetti1}. Such timescales arise from the dissipation rate of the nodal quasiparticles after photoexcitation by the pump pulse. The collective modes acting as an energy sink of photoexcited electrons are phonons and paramagnons. However the putative generation of hot spin-fluctuations would affect only the energy dissipation of electrons with excitation energy larger that the paramagnon one (which is peaked around 200 meV) \cite{Ghiringhelli} and act on a timescale faster than 20 fs. This value can be inferred from the paramagnon linewidth \cite{Ghiringhelli} and has been reported in the initial dynamics of the electrons \cite{Giannetti1}. We conclude that paramagnons are irrelevant to the dissipation of the low energy quasiparticles whereas the dynamics observed in Fig. 2 (f) only implies scattering with lattice modes: quasiparticles first emit a subset of optical phonons that are more strongly coupled to the electronic system. After roughly $\tau_1\cong 150$ fs, the excited phonon modes thermalize with the quasiparticles whereas a weaker dissipation proceeds via coupling with low energy acoustic phonons and anharmonicities. These slower processes are responsible for the cooling of the electronic system on the $\tau_2=2.5$ ps timescale. In our analysis the parameter $A_1$ can be viewed as the fraction of electronic energy density dissipated in hot optical phonons whereas $A_2=1-A_1$ is the energy resting in the electrons once the scattering with optical phonons reached detailed balance conditions. We expect that heat diffusion takes place on a timescale longer than $\tau_2$ and that can be therefore neglected.

\section{Quasiparticle relaxation across the superconducting phase transition}

Intensity difference maps as the one shown in Fig. \ref{Fig2}(d) have been measured at $F=60 \mu$J/cm$^2$ for different pump probe delays $t$ and sample temperatures $T$. The pump-probe signal $I(t)$ has been obtained by integrating the photoexcited electrons in the blue dashed square. Figure \ref{Fig3}(a) shows the extracted temporal evolution of the normalized signal $I(t)/I_{m}$. When changing temperature, the maximal intensity $I_m$ scatters randomly around the average value with errors bars of 30\% (not shown). We ascribe these uncertainties to the movement of sample position during the cooling process. We could not identify any reproducible trend of $I_m$ for $35 \tn{K}<T<150 \tn{K}$ \cite{Lanzara2}. Anyway, the relative small variations of the maximal pump-probe signal suggest that the initial energy density of excited quasiparticles depends weakly on the sample temperature. The bi-exponential fit of $I(t)/I_{m}$ and the slow component $A_2\exp(−t/\tau_2)$ are shown by solid line and colored areas, respectively.  We observed that the slow decay time does not depend on temperature and is constant in a confidence interval $\tau_2=2.5\pm0.5$ ps. Notice in Fig. \ref{Fig3}(b) that the weight of the fast component is nearly constant above $T_c$ whereas it drops when $T$ is below the critical temperature. At $T=35$ K the weight $A_1$ has almost vanished so that nodal quasiparticles are no longer able to efficiently scatter with optical phonons. This behavior correlates to a weaker dissipation rate. Despite the large error bars, the $\tau_1$ parameter in Fig. \ref{Fig3}(c) is clearly increasing when the system is cooled below the critical temperature. The trends reported in  Fig. \ref{Fig3}(b,c) have been consistently observed on three different cleaves of Bi-2212.


Our data indicate that a remnant Cooper pairing inhibit phonon scattering channels even if the photoexcitation fluence is of 60 $\mu$J/cm$^2$. On the other hand, this pump fluence is considerably larger than the minimal value necessary for the complete destruction of phase coherence. When the sample is in equilibrium conditions the phase coherence can be monitored by diamagnetic measurements or by the electrodynamics response at low THz. The latter method has been already employed by several authors to measure the inductive response of the superconducting condensate out of equilibrium \cite{Kaindl0,Kaindl1,Beyer,Averitt,Shimano0,Shimano1,Cavalleri}. Time resolved THz spectroscopy measurements on optimally doped Bi-2212 \cite{Kaindl0} have shown that  supercurrents are completely suppressed when pumping the sample with 800 nm pulses having fluence above $F_c = 11-14\mu$J/cm$^2$ \cite{Kaindl1}. These works \cite{Kaindl0,Kaindl1,Beyer,Averitt} also indicate that long-range superconductivity is rapidly destroyed and recovers on the picosecond timescale. By a direct comparison with the time resolved THz data \cite{Beyer}, we can identify a temporal window larger than 1 ps when the material does not hold superfluid density although quasiparticles do not display a strongly inelastic scattering.

\section{Comparison between Bi2212 and Bi2201}

\begin{figure} \begin{center}
\includegraphics[width=\columnwidth]{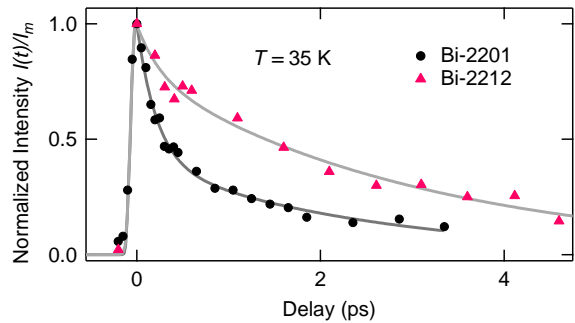}

\caption{Dynamics of photoexcited quasiparticles in Bi-2201 ($T_c = 28$ K) (circles) and Bi-2212 ($T_c = 91$ K) (triangles) acquired with pump fluence of 60 $\mu$J/cm$^2$ at the base temperature $T=35$ K. Solid lines are bi-exponential fits. The fast component $A_1$ is small only in the superconducting sample Bi-2212.}
\label{Fig4}
\end{center}
\end{figure}

The question arise if the reduction of quasiparticle dissipation persists to fluences higher than $F_c$ because of the presence of an antinodal pseudogap observed by ARPES measurements \cite{Shen,Kaminski}. This hypothesis does not explain why the dynamics in Fig. \ref{Fig3} show a clear change below  $1.3T_c \cong 120$ K, namely at the point where amplitude fluctuations of the superconducting order parameter should grow-up \cite{Alloul,Bergeal,Perfetti0}. Moreover it is inconsistent with our tr-ARPES measurements of optimally doped Bi-2201. The single layer Bi-2201 has $T_c=28$ K but develops an antinodal pseudogap already below 120 K \cite{Kaminski,He,Raffy}. Intensity maps as the ones in Fig. \ref{Fig2}(d) have been integrated in the dashed blue region. We show in Fig. \ref{Fig4} the quasiparticles dynamics upon photoexcitation fluence of 60 $\mu$J/cm$^2$ and at $T=35$K. The normalized $I(t)/I_m$ is compared to the same quantity measured on the bilayer Bi-2212. The fast component $A_1$ is clearly visible in Bi-2201 whereas it is nearly absent in Bi-2212. This comparative analysis indicates that the observed drop of quasiparticle dissipation is related to superconducting correlations but not to the pseudogap.

\section{Quasiparticle relaxation at different photoexcitation densities}

\begin{figure} \begin{center}
\includegraphics[width=\columnwidth]{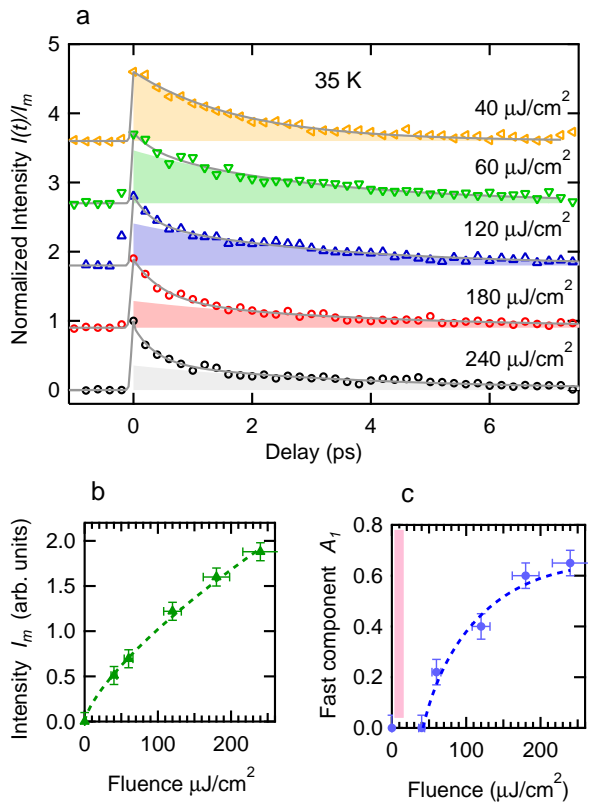}

\caption{The data of this figure have been acquired on optimally doped Bi-2212 ($T_c=91$ K) at $T=35$ K. a) Dynamics of photoexcited quasiparticles acquired at different pumping fluences. Solid lines are bi-exponential fits while the underneath colored areas stand for the slow component. The curves have been shifted by an arbitrary offset for better clarity.
b) Maximal value of the photoinduced signal $I_m$ as a function of pump fluence. The dashed line is a guide to the eye. c) Relative weight of the fast component $A_1$ as a function of pump fluence. The dashed line is a guide to the eye and the filled area is the fluence range where phase-stiffness is not totally destroyed by the optical pump pulse.}
\label{Fig5}
\end{center}
\end{figure}

Next, we set the temperature of the bilayer Bi-2212 to 35 K and we perform temporal scans increasing the pump fluence from 40 up to 240 $\mu$J/cm$^2$. Figure \ref{Fig5}(a) shows the normalized $I(t)/I_m$ and the bi-exponential fits.  In agreement with previous results \cite{Lanzara2}, we show in Fig. \ref{Fig5}(b) that $I_m$ has nearly a linear dependence on fluence for $F>F_c$. As shown by Fig. \ref{Fig5}(c), the fast scattering component $A_1$ is not detectable for pump fluence of 40 $\mu$J/cm$^2$ and grows up non-linearly at higher photoexcitation densities. An indication of this threshold has been already reported by R. Cort\'es \textit{et al.} \cite{Bovensiepen0}, who have shown that the fast quasiparticle relaxation develops for pumping fluences in the range 30-130$\mu$Joule/cm$^2$. Our data are also in agreement with the onset of the rapid dissipation channel observed at 70$\mu$Joule/cm$^2$ in transient reflectivity experiments \cite{Giannetti0}. By comparing the curve in Fig. \ref{Fig5}(c) with the measurements of M. A. Carnahan \cite{Kaindl1}, we identify a fluence regime between $14\pm3$ and $40\pm5 \mu$J/cm$^2$ when the phase coherence is lost but the fast dissipation channel is blocked.
Moreover, by comparing Fig. \ref{Fig5}(c) with Fig. \ref{Fig3}(b) we notice that the fast dissipation component $A_1$ attain similar values if: a) the sample is few tens of degrees above $T_c$ or b) the sample is at 35 K and the excitation fluence is $\sim 200 \mu$J/cm$^2$. Therefore, an excitation fluence roughly 10 times larger than the threshold $F_c$ must be employed to observe a quasiparticle dynamics that qualitative resembles to one of the high temperature phase. This finding is in strike contrast to the equilibrium state, where the superconducting fluctuations are destroyed within the Ginzburg temperature window of only $(T-T_c)/T_c\cong$ 0.3-0.4 \cite{Alloul,Ong,Perfetti0}.

\begin{figure} \begin{center}
\includegraphics[width=0.7\columnwidth]{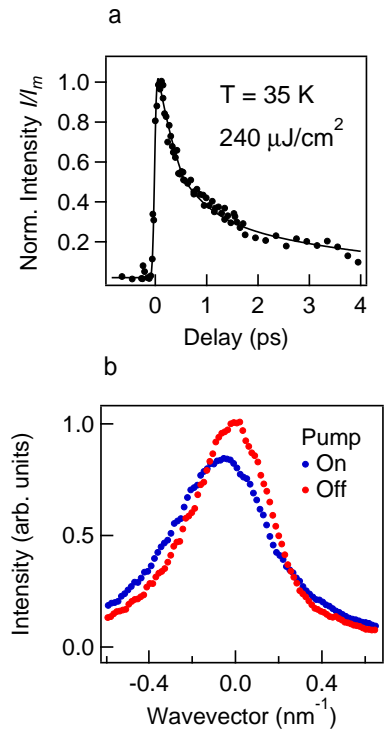}

\caption{a) Dynamics of photoexcited quasiparticles acquired on optimally doped Bi-2212 ($T_c=91$ K) at $T=35$ K and pumping fluence 280 $\mu$Joule/cm$^2$. b) Momentum distribution curves extracted by integrating the intensity maps in an energy window of 30 meV below the Fermi level. The red and blue curves stand for data acquired without pump pulse (red) and 50 fs after the arrival of a pump pulse with 120 $\mu$Joule/cm$^2$ (blue).}
\label{Fig6}
\end{center}
\end{figure}

Finally, we report in Fig. \ref{Fig6}(a) a delay scan of $I(t)/I_m$ acquired with smaller temporal steps, pump fluence of 240 $\mu$Joule/cm$^2$ and temperature of $T=35$ K. By fitting the data, we find a fast decay time $\tau_1=500$ fs, therefore consistent with the low temperature limit of Fig \ref{Fig3}(c).

\section{Discussion}

In the following we discuss our experimental findings and their implications to the physics of cuprates. 
An electron traveling ballistically with the Fermi velocity $v_F\sim 3\cdot 10^5$ m/s for an interval of time $\tau_1\sim 500$ fs covers a distance $l_m\sim 150$ nm. Therefore $l_m$ should be viewed as an upper limit of the distance explored by quasiparticles before that dissipation via optical phonon emission takes place. It is nonetheless very unlikely that electrons travel ballistically during the timescale  $\tau_1\sim 500$ fs. Instead, a random walk motion will originate from all scattering processes other than optical phonon emission.  In this case the distance explored by a quasiparticle would be smaller than $l_m$. A purely diffusive model with diffusion constant $D=l_b v_F$ gives $l=\sqrt{D\tau_1}=\sqrt{ l_b l_m}$ where $l_b<<l_m$ is the real mean free path of non-equilibrium quasiparticles. We estimate $l_b$ from an analysis of Momentum Distribution Curves (MDC) \cite{Bovensiepen1}. This procedure is not straightforward, as the FWHM of the MDCs is dominated by our finite resolution. In equilibrium and at low temperature, the intrinsic linewidth of the quasiparticle is  very small. High resolution ARPES measurements are limited by surface imperfections and provide an upper limit of 0.06 nm$^{-1}$ \cite{Shen}. Out of equilibrium conditions, the quasiparticle mean free path becomes considerably shorter. Indeed Figure \ref{Fig6}(b) shows that the photoexcited quasiparticle peak is $\cong 0.15$ nm$^{-1}$ larger than the equilibrium one. Therefore the mean free path at 50 fs after the arrival of the pump pulse is roughly $l_b\cong(0.06+0.15)^{-1}$ nm $\cong 5$ nm. This suggests that even photoexcited BSCCO is in the clean limit \cite{Tinkham}, i.e. the quasiparticles experience a mean free path $l_b$ larger than the pairing range $\xi_0 \cong 1-2$ nm \cite{Vidal}. By applying the diffusion equation to our problem we find $l=\sqrt{l_b l_m}\cong 30$ nm.

It is reasonable that an electron pairing over a distance $\xi>l$ inhibit the quasiparticles scattering. Therefore, inelastic scattering can be an effective probe of superconducting correlations on the characteristic scale $l$. Most interestingly, our experiment monitor short range correlations while the system is far from equilibrium conditions. Not much is known about the excited state generated upon irradiation with 1.5 eV. Clearly, the primary photoexcited electrons trigger a cascade of secondary processes that dephase and break Cooper pairs. Albeit long range superconductivity is lost when $F>F_c$, the short range superconducting correlations may persist up to higher pumping fluence. Therefore, the dissipation of quasiparticles would be hampered until the finite coherence (or coarsening) length $\xi$ of such superconducting correlations becomes comparable to $\l$. Our experimental observations strongly support this scenario. Since the fast component $A_1$ saturates for $F\sim 200 \mu$J/cm$^2$, the signature of short range correlations is roughly 10 times more robust than the phase stiffness \cite{Kaindl0,Kaindl1}. Notice that this state of matter only arises when exciting the system in the condensed phase. Indeed, the temperature dependence of $A_1$ in Fig. \ref{Fig3} indicates that any signature of pairing amplitude vanishes for temperatures slightly above $T_c$. The latter finding is in agreement with the small temperature window where fluctuations have been previously reported \cite{Alloul,Bergeal,Ong}. 

It is quite natural that equilibrium and non equilibrium properties of superconductors look very different: the long wavelength amplitude fluctuations are much slower than short wavelengths ones. They can be very effective in breaking superconductivity in equilibrium whereas they are still \lq\lq frozen\rq\rq~in the transient state. Therefore, the photoexcited state is dominated by short range amplitude fluctuations and related phase fluctuations. On the long timescale we may expect the formation of vortex-antivortex pairs \cite{Maniv} whereas a coarsening phenomenon characterizes the early delays. Indeed, as noticed by Giannetti \emph{et al. } \cite{Giannetti0}, the coalescence dynamics of transient reflectivity suggests the tendency of phase separation between the normal and superconducting phase. In this context, it is important to recall that a model leading to non-equilibrium first order phase transition has been discussed in the steady state by C. S. Owen and D. J. Scalapino \cite{Scalapino}.
Similar conclusions may concern also photoinduced phase transitions of conventional supercondutors \cite{Shimano0}, Charge Density Waves (CDW) materials \cite{Lee,Huber} or magnetic ordering \cite{Wang}. Hopefully, the theoretical advances in non equilibrium condensates \cite{Littlewood} and dynamical phase transitions \cite{Canovi} may provide enlightening explanations in the near future.

Finally, our data on Bi-2212 indicate that the inelastic scattering of nodal quasiparticles can be employed as a sensitive probe of superconducting correlations with short-range. The persistence of such correlations in a photoexcited state without phase stiffness is an intriguing property of non-equilibrium condensates.

This work is supported by \lq\lq Investissements d'Avenir" LabEx PALM (grant No. ANR-10-LABX-0039-PALM), by China Scholarship Council (CSC, Grant No. 201308070040)", by the EU/FP7under the contract Go Fast (Grant No. 280555), and by the R\'egion Ile-de-France through the program DIM OxyMORE. We acknowledge enlightening discussions with Claudio Giannetti and Federico Cilento on their transient reflectivity and time resolved ARPES data.

\end{document}